\documentstyle[]{article}
\begin{document}
\vspace{10 mm}
\begin{center}
\large{{\bf The Pioneer's Anomalous Doppler Drift as a Berry Phase}}
\vspace{3 mm}
\hspace{2 cm} J.L. Rosales \footnote{
Antequera $1$-$1^{0}$-H, E-28805 Alcal\'a de Henares, 
Madrid. E-mail: JoseLuis.Rosales@esp.xerox.com} \footnote{ 
To my wife Rufina and my daughter Ana}
\vspace{4 mm}
\end{center} 
\vspace{3 mm}
\begin{quote}
\begin{center}
Abstract
\end{center}
The detected anomalous frequency drift acceleration in Pioneer's radar 
data finds its explanation in a Berry phase that obtains the quantum 
state of a photon that 
propagates within an  expanding space-time. $a_{t}$ is just the
adiabatic expansion rate and 
an analogy between the effect and Foucault's 
experiment is fully suggested. In this sense, light rays play a similar r\^ole 
in the expanding space than Foucault's Pendulum 
does while determining Earth's rotation.
On the other hand, one could 
speculate about a suitable future experimental arrangement at 
"laboratory" scales 
able to measure the local cosmological expansion rate 
$\dot{R}(0)$ using the procedure  outlined in this paper. 
\end{quote}
\vspace{2 mm}
{\it{\bf{ Introduction.}}}
\hspace{2 mm}
It is not unlikely that the anomalous Doppler drift reported 
on the Pioneer's echo signals \cite{kn:Pioneer1}, \cite{kn:Pioneer2} had a cosmological origin
\cite{kn:Rosales-SanchezGomez}, \cite{kn:Pioneer3} \cite{kn:Rosales}\cite{kn:Nottale}. 
The fact that the figure of the clock acceleration, 
$a_{t}$ almost exactly coincides with that of Hubble's constant requires a 
theoretical explanation. 
By this time, however, a
commonly held opinion was that such a kind of 
coincidences can not be assigned to any real cosmological effect since the 
expansion should only affect to bigger scales than that of the solar system
(galaxies or cluster of galaxies, for instance). 
In General Relativity, as
a local theory, holds Birkhoff's theorem that tells us that in a 
homogeneous zero-pressure cosmological model we can evacuate a spherical 
region and replace the material with a compact mass $M$ at the center, 
without affecting space-time outside the region. 
The standard view is that 
derived from the Einstein and Straus Schwarzschild vacuole 
in a FLRW background \cite{kn:Einstein-Straus1},\cite{kn:Einstein-Straus2}
- which, indeed, is unstable by construction.-
The inverse of this theorem, however, does not hold, i.e., 
we can not build a global
metric from the "averaged" set of local metrics - see, for instance, \cite{kn:Ellis}. -
It means that Einstein's 
field equations can not hold on all scales simultaneously (at least we still 
do not know how could that ever be done), remaining open the
question of the possible influence of the global averaged solution on 
the compatible local metric. Of course, those 
influences could never be dynamical -for the same reason that there is
no gravitational acceleration inside  a hollow of spherical 
mass distribution.
On the other hand, the physical metric can only be determined empirically upon 
the measurement of space and time coordinates for non simultaneous events. 
This requires synchronization of remote clocks, i.e., transmissions of 
light signals and the actual reception of the reflected ray from a remote
"mirror" - 
in a way entirely similar to the determination of the spacecraft position
in Pioneer's experiment.-
Therefore, experiments set up positively, 
after the precise and beautiful Pioneer's 
orbital determination, the possibility to bring into the experimental arena 
the important question of the physical 
influence of the expanding space-time background 
on local scales. 
The ansatz for our analysis will be
\begin{equation}
ds^2= c^2 dt^2 - R(t)^2 (dr^2+ r^2 d\Omega^2) \mbox{\hspace{ 2 mm},}
\end{equation}
where, for arbitrary closed light path of total flight time $T$ 
\begin{equation}
R(0)=R(T)=1 \mbox{\hspace{ 2 mm}.}
\end{equation}
The meaning of this constraint is that of a physical re-scaling, 
i.e., the statement that the speed of light is
a local physical constant. 
Thus, from the measurement of the local speed of light, $R(t)$ is always 
in classical accordance with Birkhoff's theorem, in other words,
an observer could never, in principle, determine (locally) 
the time variation of the scale factor. 
This statement holds for the
classical theory but have to be re-examined on the light 
of the quantum theory. 

\vspace{2 mm}
{\it{\bf The quantum state of the photon.}}
\hspace{2 mm}
The emitted photon requires a representation for its state in Hilbert space
because its trajectory is not an observable. 
In other words, the photon, does not really evolve in space-time but it is 
just its quantum state what changes in time. 
Given that Eq. (2) only states that 
at the observer's position there exists no classically observable 
cosmological gravitational field, 
we have no right to determine that $R(t)\equiv 1 $ for all time at different
not observable positions of space-time - since these 
are not empirically known.- From this perspective one
would expect to obtain non local effects on the final state of a photon
that was emitted and observed in a delayed time.
The position representation for the state 
of a photon of energy $E$ and momentum 
$\vec{p}=(E/c) R(t)\vec{l}$, with $|{\vec{l}}|=1$, that propagates in the metric given by
Eq. (1) is
\begin{equation}
<r|\phi(t);R(t)>= \frac{1}{2\pi} \exp[i\Psi] \mbox{\hspace{2 mm},}
\end{equation}
where $\Psi(r,t)$ is the Eiconal given by
\begin{equation}
\Psi(r,t) = -\frac{1}{\hbar}( Et -\frac{E}{c}\vec{l}\cdot \vec{r} R(t)) 
\mbox{\hspace{2 mm}.} 
\end{equation}
Evidently, this gives for the group velocity of the waves
\begin{equation}
\dot{\vec{r}}= \frac{\partial E}{\partial \vec{p}} = \frac{c}{R} \vec{l} 
\mbox{\hspace{2 mm}.}
\end{equation}
This is consistent with the values of the speed of light obtained from 
solving $ds^2=0 $ from Equation (1). Now, 
Equation (2) imposes $\dot{r}(0)=\dot{r}(T)=c$ as required from classical theory.

\vspace{ 2 mm}
{\it{\bf The Berry Phase.}}
\hspace{2 mm}
In terms of the period of the waves $R(0)/\dot{R}(0)\gg \tau= 2\pi\hbar/E $. 
The state then, adiabatically evolves in a path of the slow parameter 
$R(t)$ space
where $R(0)=R(T)$ holds. The general formulation of 
the adiabatic evolution in parameter space was given by M. V. Berry
\cite{kn:Berry1}\cite{kn:Berry2}. The quantum state change, in a closed path, 
under the influence of the adiabatic parameter obtaining a phase 
(the "Berry phase" ) given by
\begin{equation}
<r|\phi (T);R(T)> =<r|\phi (T);R(0)>\exp[i\gamma(r,T)] \mbox{\hspace{ 2 mm} ,}
\end{equation}
where 
\begin{equation}
\gamma = i\oint dR<\phi_R|\nabla_R|\phi_R> \mbox{\hspace{ 2 mm} ,}
\end{equation}
or, if the reflection takes place at $t=T/2$,
\begin{equation}
\gamma = i(\frac{iE}{\hbar c})\{ \int_{0}^{T/2} dR \vec{l}_{+}\cdot \vec{r}+
\int_{T/2}^{T} dR \vec{l}_{-}\cdot \vec{r} \} \mbox{\hspace{ 2 mm} ,}
\end{equation}
Now, $\vec{l}_{\pm}\cdot \vec{r} = \pm r $ and the path integral gives 
\begin{equation}
\gamma = \frac{E r}{c\hbar}\{-R]_{0}^{T/2}+R]_{T/2}^{T}\}= 
\frac{2E r}{c\hbar}\{\frac{1}{2}(R(0)+R(T))-R(T/2)\}
\mbox{\hspace{ 2 mm} .}
\end{equation}
Eq. (2) taken into account we finally obtain
\begin{equation}
\gamma =\frac{2E r}{c\hbar}\{R(0)-R(T/2)\}\simeq \frac{2E r}{c\hbar}
\{R(0)-[R(0)+\frac{T}{2} \dot{R}(0)]\}= - \frac{E rT\dot{R}(0)}{c\hbar} 
\mbox{\hspace{ 2 mm} .}
\end{equation}
Which is our main result. It depends on the total flight time $T$. 
The Berry phase is a geometric object, i.e., it 
is invariant under the election of the path. This can be seen, for
instance, if we select, before observation, a light path corresponding 
to three reflections on mirrors at the 
space-time coordinates points
$P_1=(T/4, cT/4)$; $P_2=(T/2,0)$ and $P_3=(3T/4, cT/4)$
\begin{equation}
\gamma = 
i(\frac{iE}{\hbar c})\{ \int_{0}^{T/4} dR \vec{l}_{+}\cdot \vec{r}+
\int_{T/4}^{T/2} dR \vec{l}_{-}\cdot \vec{r} \} +
i(\frac{iE}{\hbar c})\{ \int_{T/2}^{3T/4} dR \vec{l}_{+}\cdot \vec{r}+
\int_{3T/4}^{T} dR \vec{l}_{-}\cdot \vec{r} \} \mbox{\hspace{ 2 mm} ,}
\end{equation}
this gives
\begin{equation}
\gamma=\frac{E r}{\hbar c}\{ -R]_{0}^{T/4}+ R]_{T/4}^{T/2}-R]_{T/2}^{3T/4}+
R]_{3T/4}^{T} \} \mbox{\hspace{ 2 mm} ,}
\end{equation}
or 
\begin{equation}
\gamma=\frac{2 E r}{\hbar c}\{ \frac{1}{2}(R(0)+R(T))-R(T/4)+R(T/2) -R(3T/4)\}
\mbox{\hspace{ 2 mm} ,}
\end{equation}
which, after (2), leads to 
\begin{equation}
\gamma\simeq\frac{2 E r}{\hbar c}\{ R(0)-
(R(0)+\dot{R}(0)\frac{T}{4})+R(0)+\dot{R}(0)\frac{T}{2} 
-(R(0)+\dot{R}(0)\frac{3T}{4})\}=
- \frac{E rT\dot{R}(0)}{c\hbar} \mbox{\hspace{ 2 mm} ,}
\end{equation}
that coincides with (10).
Of course, one could make indefinitely many additional subdivisions of the path 
without changing the result. It means that 
we could 
speculate about a suitable future experimental arrangement at 
"laboratory" scales 
able to measure the local cosmological expansion rate 
$\dot{R}(0)$ using the procedure of this example. 
One does not really need
a spacecraft as the classical apparatus (mirror) for the reflection of the photon.
Let us now define the constant with dimensions of an acceleration
\begin{equation}
a_P\equiv \dot{R}(0) c \mbox{\hspace{ 2 mm} ,}
\end{equation}
so as to write Eq. (10) as
\begin{equation}
\gamma = -( \frac{E}{c^2}) \frac{a_P r }{\hbar}T \mbox{\hspace{ 2 mm} .}
\end{equation}
Hence, the final state of the photon is given by
\begin{equation}
<r|\phi(T);R(T)> = \frac{1}{2\pi}\exp\{ -\frac{i}{\hbar}
[(E+\frac{E}{c^2}a_P r)T-\frac{E}{c}r]\} \mbox{\hspace{ 2 mm} .}
\end{equation}
Now, the measurable Hamiltonian at time $T$ is
\begin{equation}
i\hbar\frac{\partial}{\partial T} <r|\phi(T)>=H<r|\phi(T)> 
\mbox{\hspace{ 2 mm} ,}
\end{equation}
for
\begin{equation}
H(r)= E+\frac{E}{c^2}a_P r \mbox{\hspace{ 2 mm} ;}
\end{equation}
while, for the measurable
momentum we get
\begin{equation}
-i\hbar \nabla_r <r|\phi(T)>=\vec{p}<r|\phi(T)>
\mbox{\hspace{ 2 mm} ,}
\end{equation}
where
\begin{equation}
\vec{p}=(\frac{E}{c}-\frac{E}{c^2}a_P T)\vec{l} \mbox{\hspace{ 2 mm} .}
\end{equation}
\vspace{5 mm}
{\it{\bf Physical interpretation.}}
\hspace{ 2 mm}
Given that the Berry phase is geometric, no new force
is involved; nevertheless, we could try to 
{\it wrongly} translate into the language of classical physics our results. 
Let us  consider,  from Eq. (21) 
\begin{equation}
E\simeq pc(1 +\frac{a_P T}{c}) +O(\frac{a_P T}{c})^2 \mbox{\hspace{ 2 mm} ,}
\end{equation}
so that the Hamiltonian in Eq. (19) be written as
\begin{equation}
H(r,p;T)=pc(1+\frac{a_P}{c}T)+pr\frac{a_p}{c}+ O(\frac{a_P T}{c})^2 \mbox{\hspace{ 2 mm} .}
\end{equation}
We can now derive  a measurable blue shift in the frequency of the vector state
very easy; to see how, we notice that the energy changes in time according to
\begin{equation}
\dot{H}=\frac{\partial}{\partial T} H(r,p;T)=p a_P \mbox{\hspace{ 2 mm} ,}
\end{equation}
on the other hand, we get, for the momentum
\begin{equation}
\dot{p}=-\nabla_r H(r,p;T)=-p\frac{a_P}{c} \mbox{\hspace{ 2 mm} .}
\end{equation}
Now, recall $H= \hbar \omega$ and $p=\hbar k$, so that, Eq. (24) and (25) obtain 
equivalently 
\begin{equation}
\dot{\omega}=a_P k \mbox{\hspace{ 2 mm} ,}
\end{equation}
\begin{equation}
\dot{k}=-\frac{a_P}{c} k \mbox{\hspace{ 2 mm} .}
\end{equation}
That is
\begin{equation}
\dot{\omega}+c\dot{k}=0  \mbox{\hspace{ 2 mm} ,}
\end{equation}
or
\begin{equation}
\omega(T)+ck(T)=\omega(0)+ck(0) \mbox{\hspace{ 2 mm} .}
\end{equation}
Solving together Eq. (26) to Eq. (29), we get
\begin{equation}
k(T)=k(0)(1-\frac{a_P T}{c}) \mbox{\hspace{ 2 mm} ,}
\end{equation}
and, after Eq. (29) 
\begin{equation}
\omega(T)=\omega(0)+ck(0)\frac{a_P T}{c}=\omega(0)(1+a_t T) \mbox{\hspace{ 2 mm} .}
\end{equation}
Where $\omega(0)=ck(0)$ was used. This is the reported Pioneer blue shift anomaly, 
correspondig to an apparent clock acceleration given by
\begin{equation}
a_t=a_P/c= \dot{R}(0) \mbox{\hspace{ 2 mm} .}
\end{equation}
An important result can also be derived with respect to the effective velocity of light. 
Notice that 
\begin{equation}
\dot{r}=\frac{\partial}{\partial p}H(r,p;T)=c(1+\frac{a_P T}{c})+\frac{a_P}{c}r \mbox{\hspace{ 2 mm} ,}
\end{equation}
whose solution is 
\begin{equation}
\dot{r}(T)=c(1+2\frac{a_P T}{c}) \mbox{\hspace{ 2 mm} ,}
\end{equation}
obtaining 
\begin{equation}
\omega(T)=\dot{r}(T) k(T)\mbox{\hspace{ 2 mm} .}
\end{equation}
These expressions are compatible with other models that   
derive the Pioneer anomaly like Ranada's (see \cite{kn:Ranada} and references therein); 
this author claims that,  owing to 
the existence of quantum vacuum fluctuations and the cosmological expansion of space, the
velocity of light increases with time in a way analogously to the expression given above
for $\dot{r}$. Ranada uses such a kind of dependence in time  
to solve Maxwell wave equation  and derives the following compatibility condition between 
$k$, $\dot{r}(T)$ and the frequency $\omega$ (in our notation)
\begin{equation}
k=\frac{\omega}{\dot{r}}(1+\frac{\dot{\omega} T}{\omega}) \mbox{\hspace{ 2 mm} .}
\end{equation}  
We  might see that our conclussions are compatible with Ranada's empirical
interpretation of the effect
since our derived conclusions are particular solutions from his compatibility condition. 
Nonetheless, Ranada physical solution
is different from the one given here in that it implies 
that observations of the wavelength fail 
to find any effect, contrary to the  solution given in this paper that predicts 
a time variation given by
\begin{equation}
\lambda(T)=\lambda(0)(1+\frac{a_P T}{c}) \mbox{\hspace{ 2 mm} .}
\end{equation}  
\vspace{5 mm}
{\it{\bf Conclusions.}}
\hspace{ 2 mm}
 From the point of view of this paper, the "Pioneer effect" detected  in 
radar signals \cite{kn:Pioneer1}, \cite{kn:Pioneer2}, \cite{kn:Pioneer3}
should have  nothing to do with the probe but only with the fact that the spacecraft 
is acting as a "mirror" for light signals, thus, being the classical apparatus of
a quantum system that
is locally being monitored by the global expanding space-time metric.
Of course, this can only be obtained for a photon
so there is not physical acceleration of any kind. 
This important feature of the effect finally explains why planets are
not sensible to that acceleration.
The Doppler anomalous phase shift finds its explanation on a Berry phase, 
a geometrical effect. A non dynamical element of the quantum evolution in the 
expanding space background. 
This demonstrates entirely that the effect should not affect to the planets but only 
to light and that it is wrongly interpreted as a dynamical acceleration, 
being fully equivalent to a calibration effect similarly to the Foucault Pendulum 
angle defect in measuring Earth rotation 
(the Hannay's angle which, indeed, is the classical analog to the Berry quantum phase, 
also a geometrical effect). In this sense, light rays play a similar r\^ole 
in the expanding space than Foucault's Pendulum 
does while determining Earth's rotation.
On the other hand it also relates the effect to the very non-locality of quantum mechanics
since  it is just the geometry of the Universe what monitors the quantum state of the 
photon. The result has nothing to do with dynamics and, therefore, 
it does not violate Birkhoff's theorem. Moreover, 
the measurable anomaly only depends on the "Time of Flight" of the photon, since
it has also  been  demonstrated that it
does not really depends on the location of the spacecraft, for instance, 
a geostationary system of satellites would obtain the same result in case that 
the time of flight were to last enough 
(up to obtaining a similar optical resolution to that of the Pioneer experiment.)
But, this is standard physics. No new physics is involved; 
the very result has nothing to do with the probe but only with non-local properties of the quantum state.

\end{document}